# In which fields do ChatGPT scores align better than citations with research quality?

Mike Thelwall
School of Information, Journalism and Communication, University of Sheffield, UK. https://orcid.org/0000-0001-6065-205X m.a.thelwall@sheffield.ac.uk

Although citation-based indicators are widely used for research evaluation, they are not useful for recently published research, reflect only one of the three common dimensions of research quality, and have little value in some social sciences, arts and humanities. Large Language Models (LLMs) have been shown to address some of these weaknesses, with ChatGPT-4o mini showing the most promising results, although on incomplete data. This article reports by far the largest scale evaluation of ChatGPT-4o mini yet and also evaluates its larger sibling ChatGPT-4o and ChatGPT-5 mini. Based on comparisons between LLM scores, averaged over 5 repetitions, and departmental average quality scores for 107,212 UK-based refereed journal articles, ChatGPT-4o is marginally better than ChatGPT-4o mini in most of the 34 field-based Units of Assessment (UoAs) tested, although combining both gives better results than either one. ChatGPT-4o scores have a positive correlation with research quality in 33 of the 34 UoAs, with the results being statistically significant in 31. The most substantial exception is Physics, for which citations are more useful. ChatGPT-4o scores had a higher correlation with research quality than long term citation rates in 21 out of 34 UoAs and a higher correlation than short term citation rates in 26 out of 34 UoAs. ChatGPT-5 mini has even stronger correlations overall. In summary, the results give the first large scale evidence that ChatGPT-4o and ChatGPT-5 mini are competitive with citations as new research quality indicator sources.

## Introduction

Large- and small-scale research evaluations frequently use citation-based indicators to support, and in some cases replace, expert judgement. Nevertheless, they have little value in some fields and for newly published research (Wang, 2013). In addition, if they are used as indicators of research quality then primarily reflect scholarly impact, rather than rigour, originality and societal impact (Aksnes et al., 2019), which are also widely considered to be important aspects of research quality (Langfeldt et al., 2021). Moreover, even for scholarly impact, despite some citations acknowledging prior influences (Merton, 1973), others don't (MacRoberts & MacRoberts, 2018; Seglen, 1998) and the rationale for citations is that the "noise" averages out in some fields (van Raan, 1998). Alternative indicators like patent metrics (Hammarfelt, 2021) and altmetrics (Priem et al., 2012) have been proposed to partly fill these gaps but Large Language Models (LLMs) seem to have the promise of addressing a wider range of the limitations of citation-based indicators, albeit with the introduction of new problems, and one research funder is now using them in his role (Carbonell Cortés et al., 2024).

LLMs are multilayer networks of billions of nodes configured into the transformer architecture (Vaswani et al., 2017) and trained through being fed with enormous numbers of texts. Generative Pretrained Transformers (GPTs) are LLMs trained to predict likely continuations of texts fed to them (Naveed et al., 2023). The best known GPTs, like ChatGPT, have an additional training stage that involves learning to respond appropriately to human instructions or task descriptions (Ouyang et al., 2022). The different types of LLM have a wide-ranging ability to conduct natural language processing tasks, like opinion mining, with better performance than previous technologies (Min et al., 2023). In addition, ChatGPT has good performance on most standard natural language processing tasks without any of the normal customization and fine-tuning stages that are usually necessary (Kocoń et al., 2023). Recently, a large-scale study has shown that ChatGPT can give reviewer-type feedback on academic papers that the authors find useful (Liang et al., 2024), suggesting that it might also be able to score papers for quality-related dimensions.

A few previous studies have used LLMs to score academic papers for research quality, and one to predict longer term citation counts (de Winter, 2024). Two have demonstrated statistically significant positive correlations with peer review scores for submitted work for individual conferences or journals, although a limitation in both cases was that the scores were directly published on the web so the LLM may have "cheated" by knowing them in advance (Thelwall & Yaghi, 2025b; Zhou et al., 2024). Small-scale studies with private scores have also found statistically significant positive associations between ChatGPT predictions and pre-publication referee recommendations for one journal (Saad et al., 2024) and post-publication quality scores for 51 papers by one author, with similar correlations whether article full text or just the title and abstract is fed in (Thelwall, 2025ab). These, and the review feedback quality evidence (Liang et al., 2024), give reassurance that the results for datasets with public scores may not be unduly influenced by the scores being public. Three general lessons from these studies are that the best input for quality prediction purposes seems to be the title and abstract of a paper, rather than its full text (because the correlations are similar and it is cheaper/faster to feed less text), and that the default parameter settings for ChatGPT are appropriate for the task and do not need modification (Thelwall, 2025a). For this task, zeroshot seems adequate and fewshot does not bring substantial improvements (Thelwall & Mohammadi, 2025). LLMs are known to give as good results on document summaries as on full text for some tasks (Mohtadi et al., 2025), perhaps because a good summary serves to emphasize the most important aspects of a document. Other LLMs assessed for research quality scoring include Gemini (Thelwall, 2025c), Gemma3 (Thelwall, 2025e), Llama4 Scout, Magistral, Qwen3 32b, and DeepSeek R1 32b (Thelwall, 2025b), but none have clearly outperformed ChatGPT yet, even the reasoning models.

Two large scale studies have used ChatGPT to evaluate the quality of published academic journal articles, both using partly public quality scores. In both cases, the quality scores were average departmental research quality for UK-based research submitted to the national Research Evaluation Exercise (REF) in 2021. The scores for individual articles are neither published nor known, but the percentages of outputs for each department in each of the 34 broadly field-based Units of Assessment (UoAs) that attained each of the four quality levels (1*, 2*, 3* and 4*) are published in a spreadsheet and a dynamic website. It is not clear whether it is practical for ChatGPT to access and interpret this tabular format information, but it is technically possible that it did. In any case, the research quality gold standard used in both these studies (and the current paper) for each journal article was the average score of journal articles submitted to REF2021 for all articles from the same department. The first study sampled 100 articles from the top scoring departments and 100 from the bottom scoring in each UoA and obtained ChatGPT-4o mini scores for them, averaging the result from 30 repetitions of the same query. This found a positive association between ChatGPT-4o mini scores and the average research quality of the submitting department in 33 out of the 34

UoAs, with Clinical Medicine being the exception (Thelwall & Yaghi, 2025a). A follow-up study widened the sample to all articles without short abstracts in the anomaly, Clinical Medicine, and found an overall positive correlation for both ChatGPT-4o and Chat GPT 4o-mini (Thelwall et al., 2025). Whilst, in theory, the combination of these two papers suggests that ChatGPT quality scores positively associate with expert research quality judgments in all fields, it is also possible that the method used to overturn the Clinical Medicine anomaly would do the same for some of the other fields. In other words, it is possible that the sampling method used in the first paper would mask a negative correlation in some of the remaining 33 UoAs.

The current paper addresses the following research gaps. First, it compares ChatGPT-4o mini scores, as used in most prior research, with ChatGPT-4o scores, the more powerful version of the LLM. ChatGPT-4o mini is a smaller, cheaper version of ChatGPT-4o. Although the nature of the difference between them is proprietary knowledge, one way of simplifying a LLM is quantisation, which involves lowering the accuracy of the parameters in the model (Lin et al., 2024), but it may also not be derivative but instead have a simpler architecture, perhaps with the same training data and design philosophy. Second, it replicates the prior study with ChatGPT-4o mini scores for samples of 200 articles from each UoA (Thelwall & Yaghi, 2025a), except expanding to complete coverage of each UoA, at least for articles without short abstracts. Third, it compares the strength of ChatGPT-4o scores with shorter term and longer-term citation-based indicators across fields. This last point is important to help assess the relative value of the two indictor sources. For example, REF2021 must use short-term citation-based indicators to fit in with the evaluation timescale but other evaluations may have more time. ChatGPT-4o rather than ChatGPT-4o mini is the focus of RQs 2-4 on the basis that prior research has found it to be superior for this type of task (Thelwall & Yaghi, 2025a). Fourth, ChatGPT-5 was released whilst this paper was under review, so an additional research question addresses the potential that it gives improved performance.

- RQ1: Do ChatGPT-4o research quality scores correlate more strongly than ChatGPT-4o mini research quality scores with expert quality scores for journal articles? RQ1a: Does averaging scores from both give better results?
- RQ2: Do ChatGPT-4o research quality scores correlate positively with research quality scores in all fields (UoAs in this case)?
- RQ3: Do ChatGPT-4o research quality scores correlate more positively than citation-based indicators with research quality scores for short term citations?
- RQ4: Do ChatGPT-4o research quality scores correlate more positively than citation-based indicators with research quality scores for medium term citations?
- RQ5: Is ChatGPT-5 an improvement on ChatGPT-4o for this task?

# Methods

The research design was to obtain a large set of journal articles with proxy quality scores in all fields and then to correlate these with quality scores for them from ChatGPT-4o, ChatGPT-4o mini and ChatGPT-5 mini and with citation-based indicators.

## *Data: REF2021 journal articles*

The raw data used was the set of journal articles submitted to the UK REF2021 national evaluation. This involved 157 publicly funded research institutions (mainly universities) submitting their best outputs from 2014 to 2020 for systematic expert quality evaluation (REF, 2019; Sivertsen, 2017). As part of this, each submitted article was given a single overall research quality score for originality, significance and rigour by two field experts, from a group of 900 field experts, supported by 220 research users. REF2021 involved assessments of 185,594 research outputs (including duplicates) (REF, 2021), but only the journal articles are included here, for consistency. A complete list of the outputs (but not the scores) is available online (results2021.ref.ac.uk/outputs). The scoring process took a year to complete and the destination of about £14 billion in block grants depended on the outcomes, so it was probably taken seriously by the evaluators. The scoring system is as follows, excluding the very rare 0 category for research that is out of scope or low quality (REF, 2022):

- **Four star**: Quality that is world-leading in terms of originality, significance and rigour.
- **Three star**: Quality that is internationally excellent in terms of originality, significance and rigour but which falls short of the highest standards of excellence.
- **Two sta**r: Quality that is recognised internationally in terms of originality, significance and rigour.
- **One star**: Quality that is recognised nationally in terms of originality, significance and rigour.

The scores allocated to outputs have been destroyed to prevent individual scores from being known but the number of outputs achieving each quality level is published online for each UoA and each “submission” (results2021.ref.ac.uk). Here a submission is a collection of outputs submitted to a single UoA by a single higher education institution. This can intuitively be thought of as the work of a university department (the terminology used in the remainder of this paper), although there are many other organisation names and structures, and outputs will often be combined from multiple organisational units into a single submission. The departmental average scores can naturally be calculated as the average of the star levels, weighted by the percentage of submissions achieving that level. This would be a reasonable approach but would be imperfect for the current article because some outputs are not journal articles. For the current study, prior private access to the individual scores for the journal articles had allowed departmental averages to be calculated for the journal articles alone before the mandatory data deletion date, and these averages were used instead.

Combining the Microsoft Excel spreadsheet of output names with the departmental average scores gives the gold standard research quality data used in this paper: REF2021 journal articles and the average quality of the journal articles from the submitting department. This latter serves as a proxy for the unavailable individual article quality scores. Although this indirect approach is not ideal, there is no other large scale source of research quality scores for all fields. Moreover, in the absence of departmental biases, higher correlations with the proxy (departmental) scores, correspond to higher correlations with the underlying article quality scores. Here, the departmental averaging has a dampening effect, with the effect being most pronounced for UoAs for which the differences between departmental average scores differed the least and for which within-submission quality scores varied the most.

For the current study, the prior private access to the individual article scores had also been used to calculate the average correlation between the article level scores and the departmental average scores. In the absence of departmental biases, these correlations are the theoretical maximum correlation for any indicator with the departmental averages. For example, if an indicator was 100% accurate then its correlation with the departmental average would be the same as the expert score correlation with the departmental average. Of course, experts make mistakes and so 100% agreement is impossible in practice.

Nevertheless, these theoretical maximums were used as benchmarks for the indicator correlations, as described below.

Articles with short abstracts were removed from the dataset because these tended to represent articles without abstracts, with trivial abstracts, or errors, and so LLMs seem unlikely to be able to score them well. For each UoA, the articles with the shortest 10% of abstracts were removed for this reason. The 10% was determined heuristically from an examination of the data before submitting the results to ChatGPT. A round number (10%) was chosen to reduce the risk of tailoring the sample size to any particular article size as a cut-off. This approach risks biasing the results if the genuine short article abstracts behave differently to long article abstracts in regard to the correlation with quality scores. The alternatives would either have been to keep the short articles, which would have introduced errors through accidentally truncated abstracts and short form articles, or to have checked and excluded the articles with short abstracts individually. The latter approach was impractical in this multidisciplinary collection because it can be difficult to judge whether individual articles are short. For example, some medical journals publish short summary papers with links to full texts elsewhere and it is not clear whether they should be regarded as short or long.

## *ChatGPT-4o and 4o mini scores*

ChatGPT was chosen rather than any other LLM because it is one of the largest and most successful in text processing tasks and has been used in almost all previous research quality scoring experiments, with no other model producing better results. It would nevertheless have been helpful to compare its results against those of other LLMs but this was not necessary for the research questions, and it would have been prohibitively expensive, either for API calls or for local computing time to run the full dataset five times through any except the smallest quantised LLM.

The articles were submitted to ChatGPT-4o and ChatGPT-4o mini through the Applications Programming Interface (batch mode) with system instructions derived from the REF2021 evaluator instructions (REF, 2019), as previously used (Thelwall & Yaghi, 2025a). As shown in the Appendix, these instructions define the quality levels and explain that the dimensions of quality to be assessed are rigour, originality, and significance. They also loosely define these dimensions. The main prompt given was, "Score this article:" followed by the article title, a new line, the word "Abstract", another new line, and the article abstract as a single line.

Article titles and abstracts were used for multiple reasons. First, prior research has found scores from titles and abstracts to be better than scores from full texts for ChatGPT-4o (Thelwall, 2025a). Second, full texts were not available for the full dataset. Third, ingesting full texts increases the cost for each query and this would have exceeded the project budget. Fourth, full text submission may be too expensive for practical applications both because of the API call costs and the costs associated with obtaining full texts. Fifth, although submitting copyright documents to ChatGPT via its API for research purposes is legal in the UK, it is not clear whether it is legal for applications.

Each article was submitted five times non-consecutively to ChatGPT-4o and 4o-mini and the average of the five scores used in both cases. ChatGPT tends to give very different results for identical queries and averaging multiple outputs is a way of leveraging deeper information about the probability of different scores emerging from the model. This has been shown to substantially improve the value of the scores (Thelwall, 2025a). Five submissions was a cost-based compromise, given that more repetitions would have given better results, but each new repetition contributes less added value than the one before (Thelwall, 2025a).

The ChatGPT reports produced are narrative evaluations of the quality of the input article, almost always accompanied by either an overall score or separate scores for originality, rigour, and significance. These scores were extracted by a program written for this purpose (github.com/MikeThelwall/Webometric_Analyst, AI menu). Overall scores were used, when identified, otherwise the average of the scores for the three dimensions was calculated. When no score could be found by the rules, the author was prompted to identify the score or flag that no score had been given. All the reports read by the author were plausible and internally consistent (e.g., never giving scores of 2* for each quality dimension but 3* overall and with a single observed exception always giving the overall score that was closest to the average of the three scores, if a whole number).

## *Citation-based indicators*

To obtain citation-based indicators for the articles, the REF2021 articles were matched to Scopus records either by DOI or, if no DOI match existed, by title, with the author checking the match for accuracy. This matching was reused from a previous project with a copy of Scopus downloaded in January-March 2021.

The citation-based indicator used was the Normalised Log-transformed Citation Score (NLCS) (Thelwall et al., 2023). This is a field and year normalised indicator. It was used in preference to the more common Normalised Citation Score (NCS) (Waltman & van Eck, 2013) because it reduces skewing in the citation data, making the results less influenced by individual highly cited articles and more statistically justifiable. For this, first all citation counts c were replaced with ln(1+c) to reduce skewing (the log part of NLCS). Then, the average of these values was calculated for each narrow field and year in Scopus (using its All Science Journal Classifications system). Finally, each ln(1+c) value was divided by the average for its field and year. When an article was in multiple fields and years, these were averaged before the division.

The NLCS formula serves two purposes. First, it makes comparisons between years fairer because NLCS values are normalised for publication year. Second, it makes comparisons between fields fairer, assuming that highly cited field have an unfair advantage over less cited fields. Moreover, an NLCS of 1 indicates that an article is averagely cited for its field and year, irrespective of that field and year. Thus, it is reasonable to process NLCS values for articles from different fields and years together.

The 2021 Scopus data represents short term citations in the sense that citation counts from early 2021 would have had between 0 and 6 years to accumulate since the REF2021 articles were published between 2014 and 2020. This period is contemporary with the citation indicators that the REF2021 evaluators would have had access to during their evaluations (although it probably played a minor role in their decisions in the few UoAs that used them, if the situation did not change since the previous REF: Wilsdson et al., 2015). The same process was repeated with a recent copy of Scopus from December 2024 to represent medium term citations: almost all articles having at least 4 years of citations.

## *Data set 2: ChatGPT-5 mini scores*

Whilst this article was under review (submitted 6 April 2025; first round of reviews received 3 February 2026), ChatGPT-5 was released (7 August 2025). To keep the paper current, the articles were therefore also scored with ChatGPT-5 mini, the results compared with the ChatGPT-4o results, and the manuscript adjusted to reflect this change (e.g., removing "4o" from its title and adding RQ5). Only ChatGPT-5 mini was used and not any full version because the ChatGPT-4o results showed that there was little difference between full and mini versions of ChatGPT for this task and the full version of ChatGPT-5 would exceed the project's budget. As the results show, this was sufficient to give a positive answer to the fifth research question.

Two new prompts were used for ChatGPT-5, which had been developed through testing on datasets for another project since the original paper was submitted: one for reduced cost and one for increased accuracy. The reduced cost prompt was to ask for the score alone but encouraging fractional scores because this gives better results (Thelwall, 2025b), as follows.

> Give a score for the following article in the range 1* to 4*, including fractions. No words. No justification. No report. Just the score.
> ###
> [article title and abstract]

This is a "cheap" prompt because most of the cost of the previous ChatGPT requests was in writing the reports. Previous testing had shown that asking for a score alone gave results that were only slightly worse at a reduced cost to the prompts used for ChatGPT-4o in this paper (Thelwall & Yang, 2025).

For the score-only prompt, the same system instructions were used as before except that the parts referring to report writing were removed: "alongside detailed reasons for each criterion", ", ensuring robust analysis and feedback", and "You will maintain a scholarly tone, offering constructive criticism and specific insights into how the work aligns with or diverges from established quality levels.". Also, "You will emphasize scientific rigour, contribution to knowledge, and applicability in various sectors, providing comprehensive evaluations and detailed explanations for your scoring" was changed to, "You will assess scientific rigour, contribution to knowledge, and applicability in various sectors". This prompt was submitted five times to assess whether averaging scores would increase the value of the estimates.

The second prompt requested a structured score report, modelled on that used by the University of Sheffield to carry out this type of assessment. It requires a discussion of the three core REF quality dimensions of rigour, originality, and significance before giving a score for them and then requesting an overall report and score. This has several advantages over the prompt used for ChatGPT-4o: it closely mirrors a human task, so ChatGPT may have trained on similar tasks; it requires a discussion of the core quality dimensions before giving a score, ensuring that deliberation has the potential to influence the score, and the scores are simple to extract from the reports, ensuring error free results. The disadvantage is that the prompt is much more expensive the score-only prompt.

> Score this article in the range 1* to 4*, including fractions. Use the following format:
> Rigour report:
> Rigour score:
> Originality report:
> Originality score:
> Significance report:
> Significance score:
> Overall report:
> Overall score:
> ###
> [article title and abstract]

This prompt was only able to be submitted once, due to its cost.

The scores from both forms of ChatGPT-5 mini prompt were extracted with simple Python pattern-matching programs rather than the complex pattern matching approach used for the ChatGPT-4o responses.

### *Analysis*

The ChatGPT scores and citation rates were compared against the gold standard of submitting department average quality scores using Spearman correlation to assess the extent to which their ranks align. Bootstrapping was used to calculate 95% confidence intervals. Spearman correlations were used instead of Pearson correlations because the rank is more important than fit to a linear trend. A correlation was judged to be statistically significant if its 95% confidence interval excluded 0, which is equivalent to a standard hypothesis test. Two correlations can be reasonably safely assumed to be statistically significantly different if their 95% confidence intervals do not overlap. Even though a small amount of confidence interval overlap can occur when two values are statistically significantly different, it is safer to look for this, given that there are many potential comparisons that can be made from a single graph, increasing the chance of incorrectly concluding that two underlying correlations are different.

Previous studies have shown that ChatGPT tends to give higher scores than human experts (Thelwall, 2025a) so their key property is the rank correlation with expert scores, and their accuracy is irrelevant. This puts them on the same footing as citation-based indicators in the sense of neither giving numbers that should be interpreted as estimates of the "correct" human score. Thus, traditional machine learning metrics like precision, recall, F-measure and mean squared error are unnecessary.

A range of hybrid indicators are included in the results, such as one averaging the ChatGPT-4o and ChatGPT-4o mini scores, even though this does not directly address a research question. This is provided as additional information in case it suggests that combining different models may be a useful strategy.

Since the gold standard uses a proxy indictor of article quality in the form of departmental average journal article quality, the results are sometimes presented scaled by dividing by the correlation between the REF2021 expert scores and the departmental averages for journal articles (as mentioned above), which is the theoretical maximum.

## Results

### *RQ1: ChatGPT-4o vs. ChatGPT-4o mini*

ChatGPT-4o has a higher correlation than ChatGPT-4o mini with articles' departmental average REF2021 scores in 27 out of 34 UoAs (79%) and overall, although the differences are typically small (Figure 1). For all articles combined, the Spearman correlation between the average of the ChatGPT-4o scores and the article departmental averages is 0.420, for ChatGPT-4o mini it is slightly lower at 0.416, but if both are combined, the correlation is highest at 0.444, presumably because of the additional repetitions for averaging. It is therefore clear that whilst ChatGPT-4o is better for this task than the simpler version, both give very similar results in all fields. Moreover, since combining both improves the results consistently, they act in a supporting rather than a conflicting way.

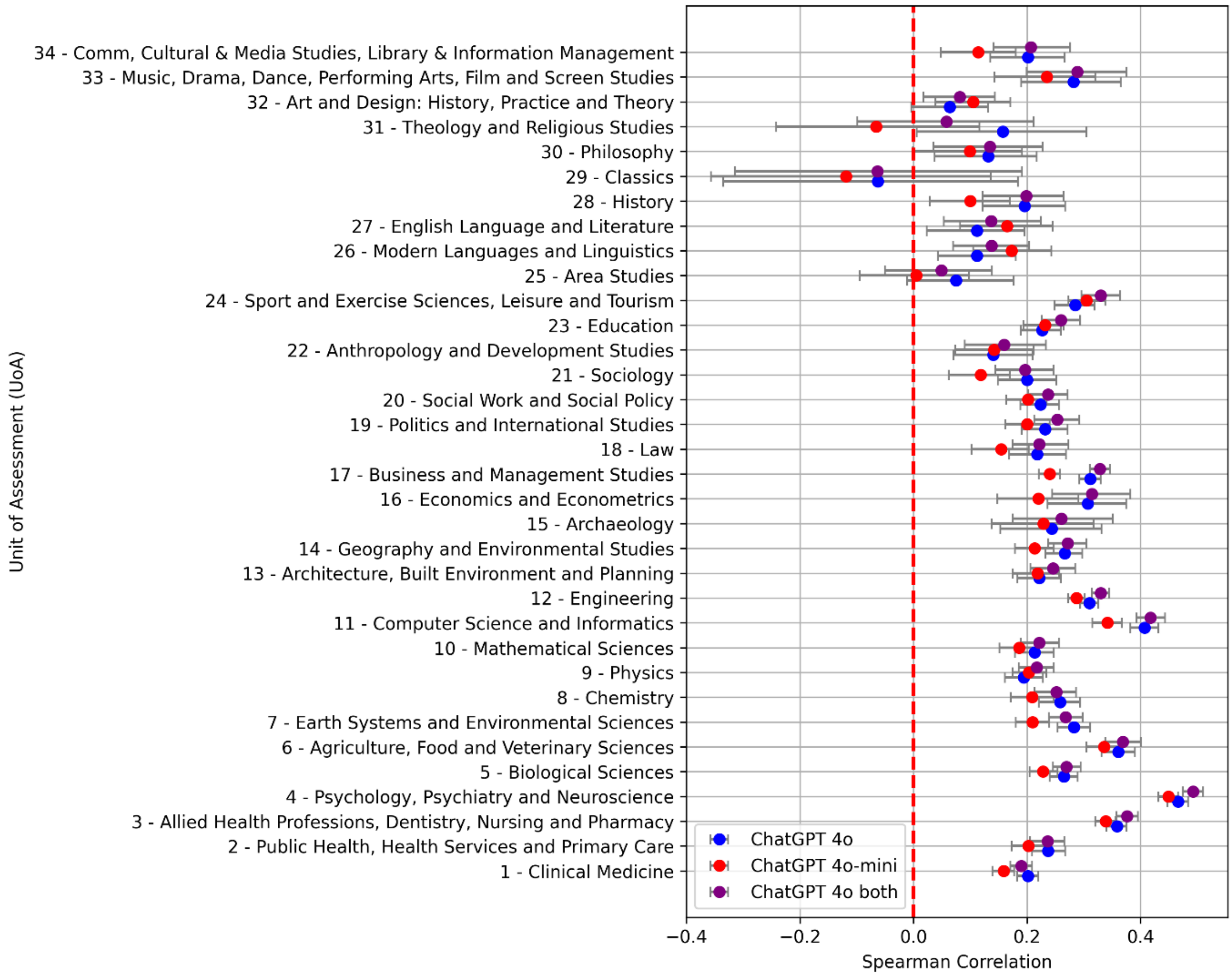

Figure 1. Spearman correlation between ChatGPT-4o scores for REF2021 articles without short abstracts and the article departmental REF2021 scores by UoA. The same for ChatGPT-4o mini. In both cases, the scores are the average of 5 independent scores (n=107,212 articles; excluding one article without a valid 4o-mini score). Error bars show 95% confidence intervals.

## *RQ2: Do ChatGPT-4o research quality indicators*

Focusing on the ChatGPT-4o scores (average of five repetitions), their correlations with the gold standard are positive in all UoAs except Classics and the positive correlation is statistically significant in 31 of the 34 fields (Figure 1: 31 of the ChatGPT-4o confidence intervals do not contain 0). Moreover, for all UoAs the 95% confidence intervals comfortably include positive correlations. Thus, there is statistical evidence that ChatGPT-4o scores correlate positively with a research quality proxy (departmental average quality) for almost all fields, and the results do not rule out the possibility that the underlying correlations are positive for all fields.

## *RQ3: ChatGPT-4o compared to short term citations*

For citation data from 2021, contemporary with the REF2021 evaluations, NLCS have a stronger Spearman correlation than ChatGPT-4o with article departmental average REF2021 quality scores in only 8 out of 34 UoAs (24%) (Figure 2). For all articles combined, the ChatGPT-4o correlation was 0.420 and the NLCS correlation was much lower at 0.223. For the UoAs with a higher early citation indicator correlation, the difference is typically small except for Classics, possibly due to a small sample size and a wide confidence interval. Early citations only have a statistically significant advantage, at least in the simplistic sense of non-overlapping confidence intervals, for Physics. Conversely, by the same standards, ChatGPT-4o scores have a statistically significant advantage in nine (UoAs 3, 4, 5, 6, 11, 12, 13, 17, 24). Thus, ChatGPT-4o has a clear advantage over citation rates based on early citations, at least for this indicator.

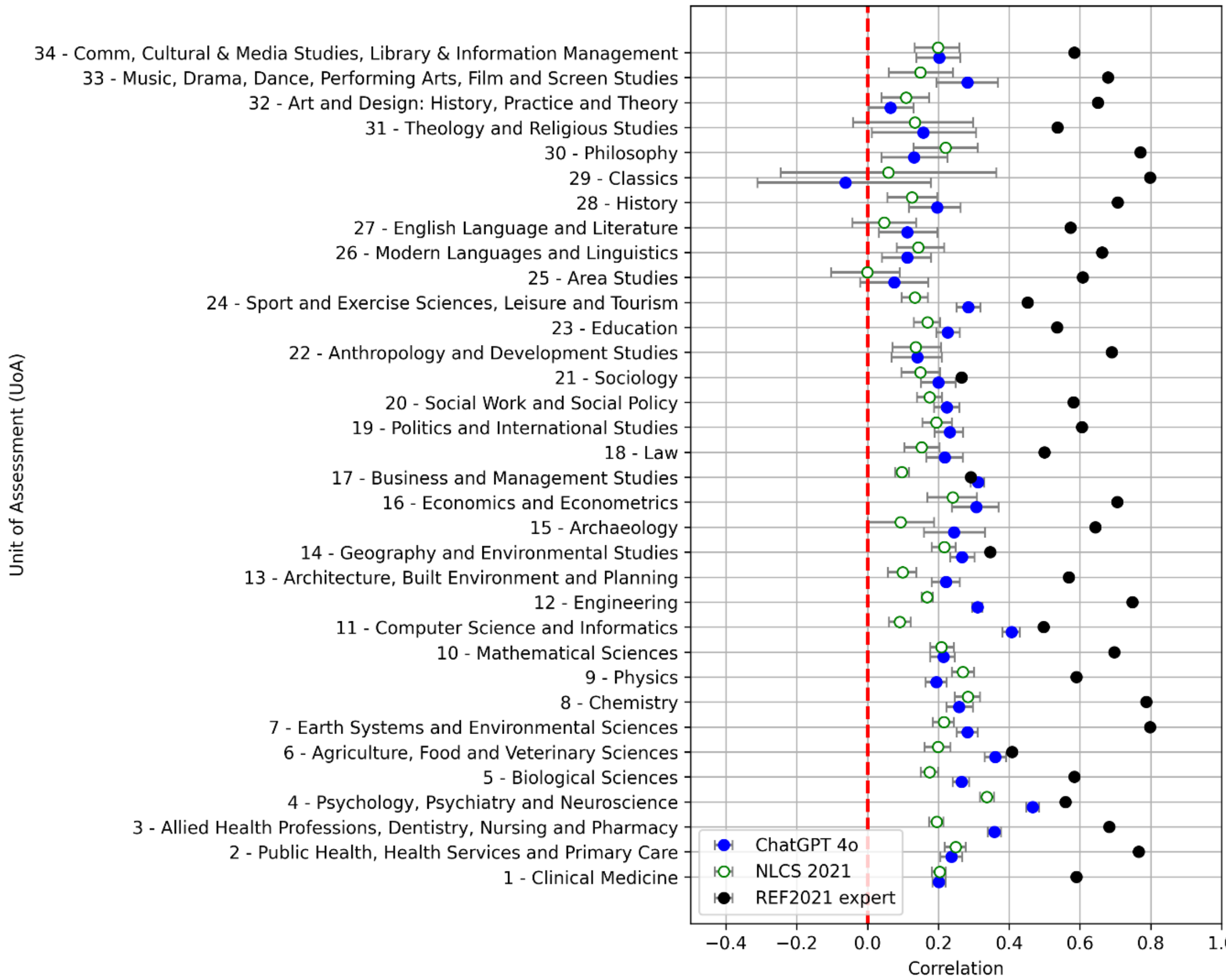

Figure 2. Spearman correlation between ChatGPT-4o scores for REF2021 articles without short abstracts and the article departmental REF2021 scores by UoA. The same for NLCS citation rates for citation data from January-March 2021 (n=107,213 articles). The black dots indicate the average correlation between REF2021 scores and departmental average REF2021 scores for all REF2021 articles.

Compared to the theoretical maximum correlation for each UoA due to the replacement of individual article quality scores with their departmental averages, the correlations range from weak to strong (Figure 3). Since one correlation ratio exceeds 1, this shows that the assumption of a lack of departmental bias in the results is untrue, at least in this one case. Here the term “bias” does not necessarily mean “unfair”. For example, it is possible that ChatGPT is better at assessing the quality of the department producing an article than the quality of individual articles. This could occur because some high-quality departments have specialities that ChatGPT scores highly irrespective of the quality of individual articles (or vice versa). Bias in the neutral sense is also suggested by the correlations varying substantially between broadly similar UoAs. Thus, this graph should be interpreted particularly cautiously.

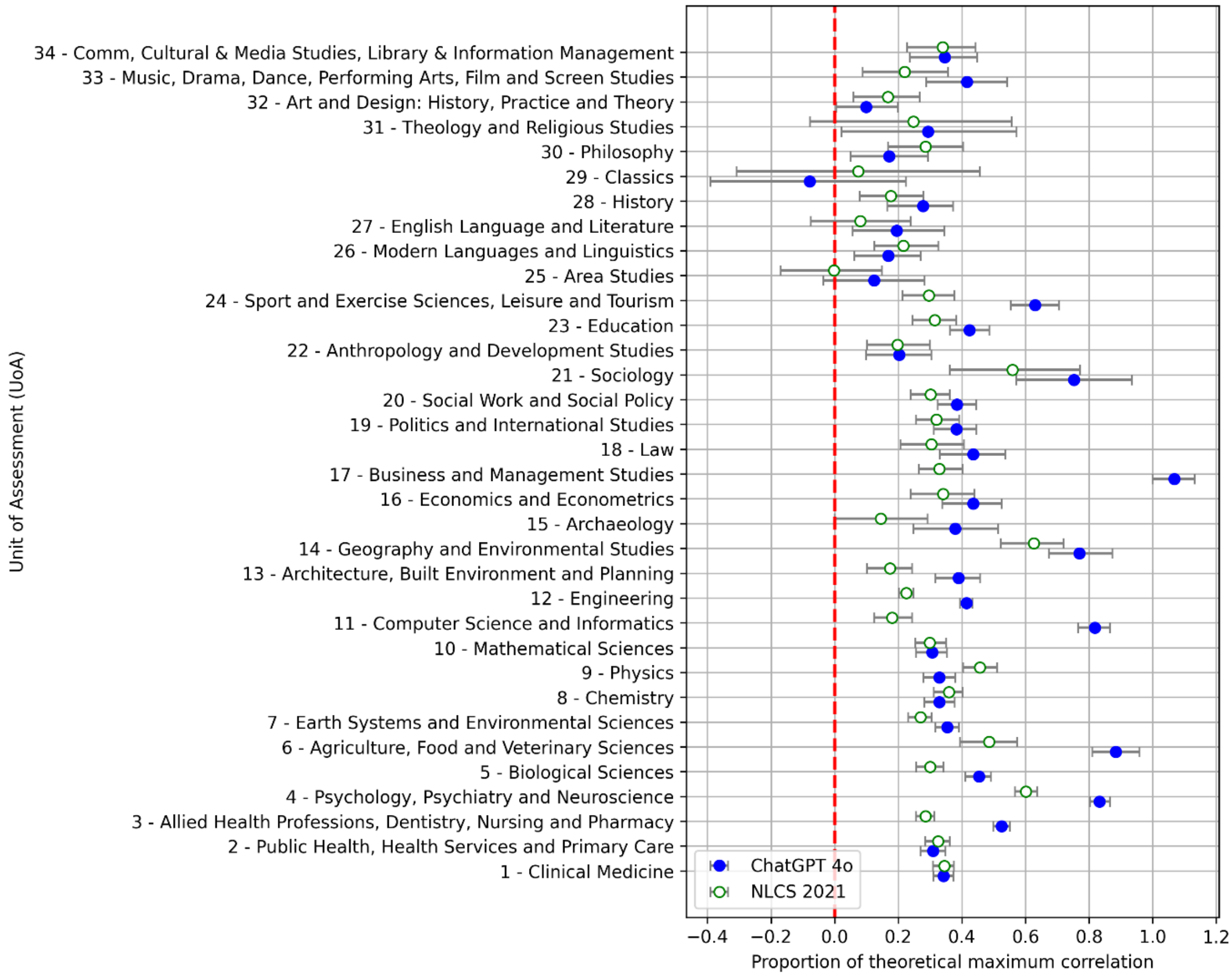

Figure 3. As above but scaled to set REF2021 expert = 1. In other words, in each field

## *RQ4: ChatGPT-4o compared to medium term citations*

For citation data from 2024, giving almost all articles at least four full years of citation data, NLCS citations have a stronger Spearman correlation than ChatGPT-4o with article departmental average REF2021 quality scores in a minority of 13 out of 34 UoAs (38%) (Figure 4, Figure 5). Overall, the ChatGPT-4o correlation was 0.419 (slightly different from the NLCS 2021 comparison due to excluding articles without 2024 data) and the NLCS correlation was still much lower at 0.227. Using the non-overlapping confidence interval simplification, the NLCS advantage over ChatGPT-4o is again statistically significant only for Physics, whereas the ChatGPT-4o advantage is statistically significant for the same nine UoAs as before (3, 4, 5, 6, 11, 12, 13, 17, 24). Thus, even with the advantage of a substantial citation window, ChatGPT-4o scores (average of five repetitions) are still better overall than citation rates, at least for the NLCS indicator.

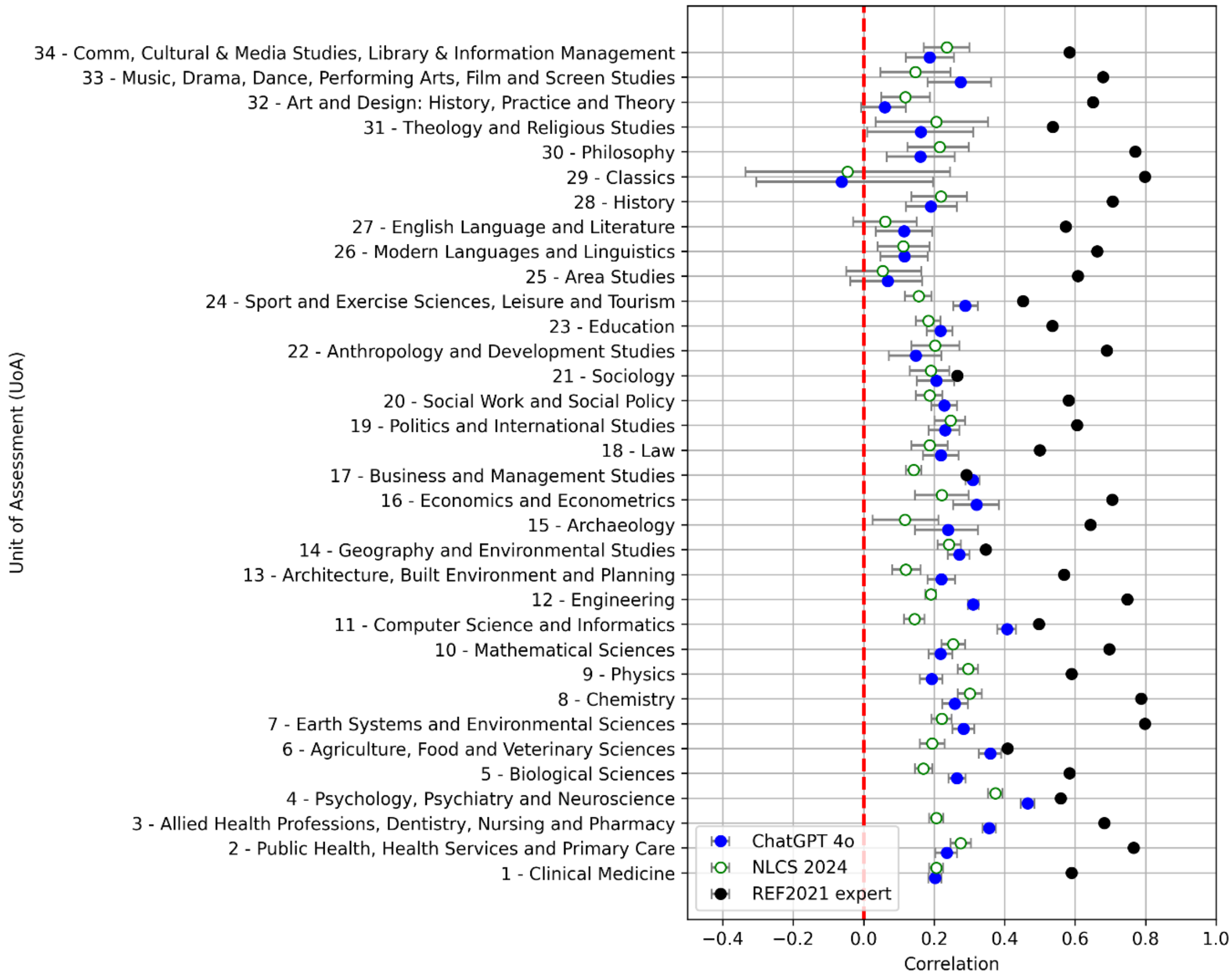

Figure 4. Spearman correlation between ChatGPT-4o scores for REF2021 articles without short abstracts and the article departmental REF2021 scores by UoA. The same for NLCS citation rates for citation data from December 2024 (n=105,049 articles; this excludes articles without a matching Scopus ID in 2024). The black dots indicate the average correlation between REF2021 scores and departmental average REF2021 scores for all REF2021 articles.

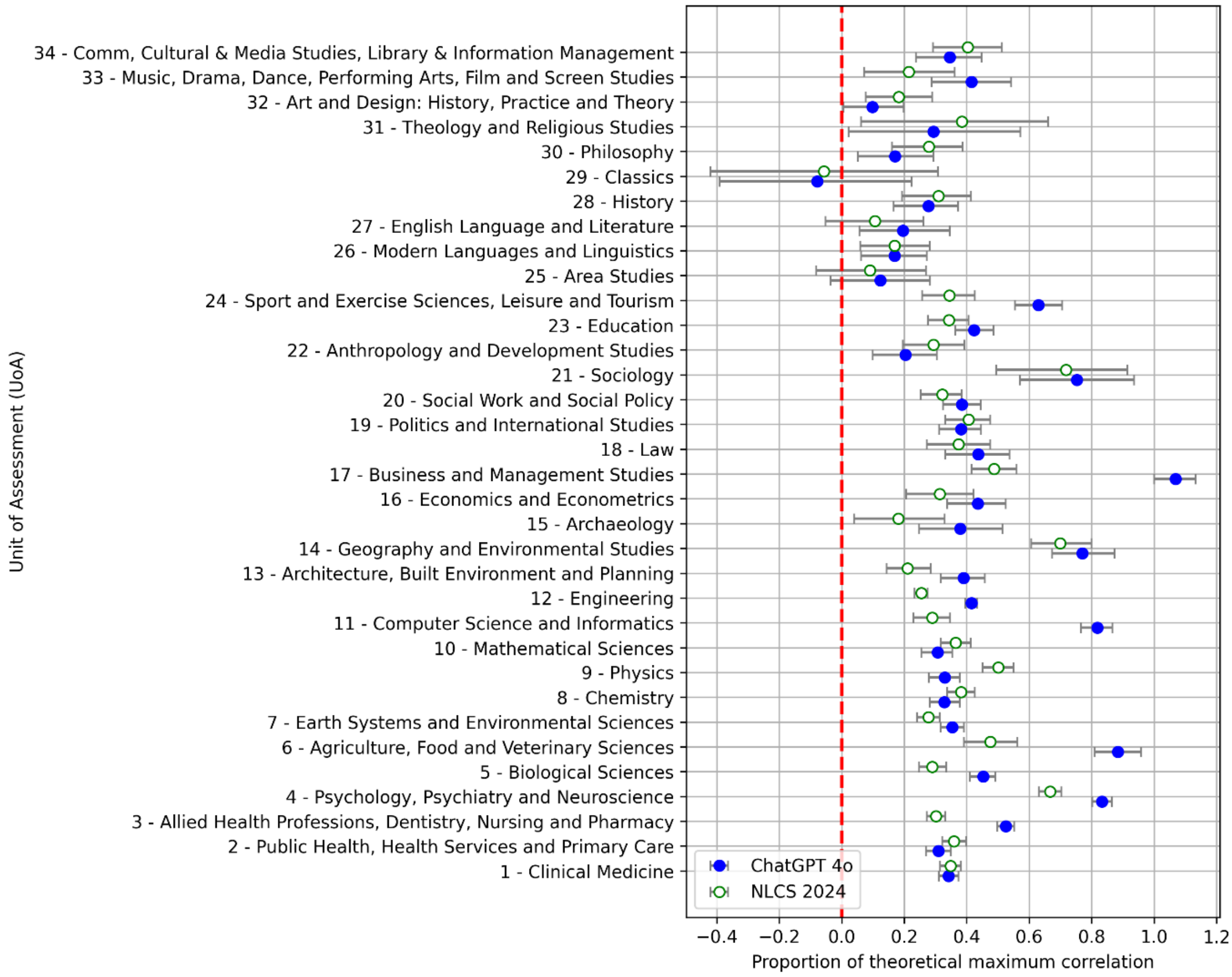

Figure 5. As above but scaled to set REF2021 expert = 1 (i.e., dividing by the correlation between the REF2021 expert scores and the departmental averages for journal articles, which is the theoretical maximum correlation).

## *RQ5: Is ChatGPT-5 an improvement on ChatGPT-4o for this task?*

For all four REF2021 main panels, both new sets of ChatGPT-5 mini scores gave substantially stronger correlations than all the ChatGPT-4o scores, despite being the smaller version of the latest ChatGPT-5 model (Figure 6). Both versions of the ChatGPT-5 mini prompts gave similar overall correlations, but the structured prompt gave noticeably stronger correlations when only single iterations are compared. The correlations were universally positive except in the Arts and Humanities (Panel D). Note that for the structured prompt, the separate rigour, originality, and significance scores did not give higher correlations than the overall score.

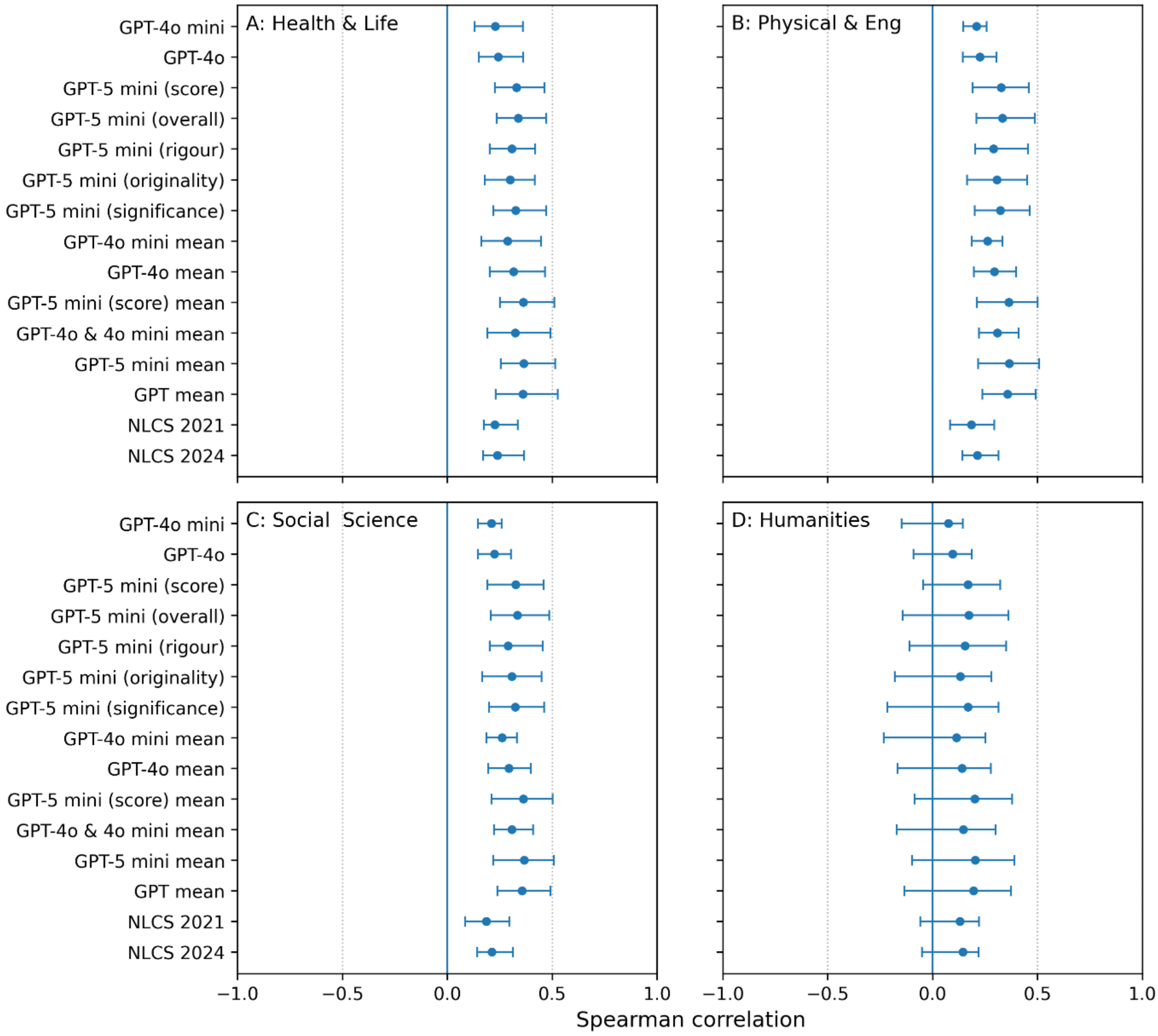

Figure 6. Sample size weighted average Spearman correlation between citation-based indicators or ChatGPT-based indicators and departmental average REF scores for articles for each of the four main panels. The correlations for indicators without (mean) are for single prompts (average correlation if multiple identical prompts were submitted), whereas the indicators with mean in their name are the average of five (or more, if averaging across multiple prompt types or models). GPT mean is the average of all scores from all prompts from both versions of ChatGPT. Error bars show the lowest and highest correlation for individual UoAs.

Comparing ChatGPT-5 mini score correlations with citation rate indicator correlations separately across the 34 UoAs (Figure 7), it gives a stronger correlation in 79% of cases (27 out of 34). Of the exceptions (Anthropology & Development Studies; Classics; History; Philosophy; Physics; Public Health, Health Services & Primary Care; Theology & Religious Studies), UoA 9 Physics is the only one for which the confidence intervals do not overlap. If only early citations are considered, then ChatGPT-5 mini is stronger in 91% of cases (31 out of 34), with the exceptions being Classics, Philosophy, and Physics.

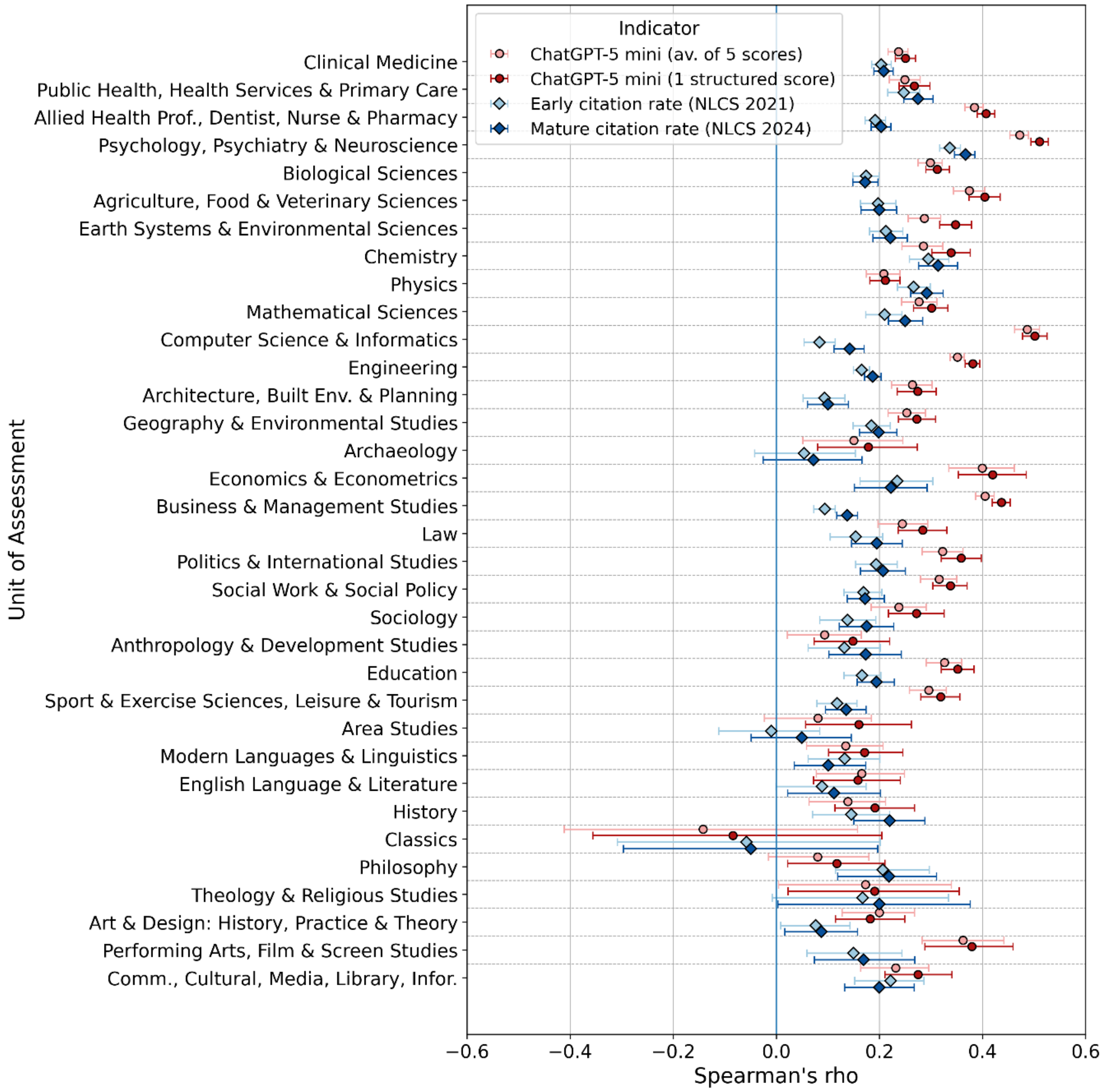

Figure 7. Spearman correlations between ChatGPT-5 mini (both prompt types) and citation rates and departmental average REF scores for articles by Unit of Assessment.

## Discussion

The most important limitation of this study is that the gold standard scores used are public, so it is possible that the positive correlations are due to ChatGPT picking up indirect signals of research quality from ingesting and connecting relevant pages from the two sections of the REF2021 website. It is not clear whether ChatGPT could do this because the connections between articles and departments are in a spreadsheet (university name and UoA name), and the connections between departments and score profiles are in a dynamic website amongst other information, so it seems unlikely that the connection between the two could be made by ChatGPT, even weakly, although it is not impossible. The likelihood of this kind of “cheating” is also reduced by evidence that ChatGPT rarely answers basic questions about journal articles, such as their publishing journal or year, when fed with a title and abstract (Thelwall, 2026), so associating more complex REF information seems unlikely. In support of the existence of an underlying connection between research quality and ChatGPT scores, prior studies have found positive correlations for private scores (e.g., Saad et al., 2024), one of which is a stronger correlation for the corresponding UoA (34) than the one found here (Thelwall, 2025a). Moreover, the input data did not include any information necessary to directly connect an article to a UoA or university (only the title and abstract) and the ChatGPT reports read only discussed the content of articles, without ever mentioning anything about the publishing journal or the authors, departments, UoAs, or universities.

A wider limitation is that the articles are from a single English-speaking Global North country and it is not clear how well they would transfer to other contexts. ChatGPT may need radically different system instructions to align to Global South concepts of research quality (e.g., Barrere, 2020) and may not be able to. Finally, the results excluded articles without abstracts or with short abstracts and it seems likely that ChatGPT would perform worse for these.

An additional consideration is that the ChatGPT correlations can be expected to be stronger if averaged over more repetitions. Thus, the relative potential strength of ChatGPT-based indicators compared to citation-based indicators may be stronger than reported here. This refers to the potential to devise a ChatGPT-based strategy to obtain stronger indicators rather than the inherent strength of ChatGPT in answering prompts. Conversely, other citation-based indicators may have a stronger correlation with research quality than does the NLCS, also changing the balance of relative strengths between the two types.

A practical limitation for the use of ChatGPT is that it is not cheap. In addition to the human time needed to extract and interpret the scores, the ChatGPT-4o queries used in the current paper cost about $3,000, with ChatGPT-4o costing ten times as much as ChatGPT-4o miniChatGPT-4o mini and the repetitions needed to improve accuracy multiplying the cost by five times. In this context, the results suggest that, for the same level of correlation with quality scores, it is cheaper to average multiple iterations of ChatGPT-4o mini than ChatGPT-4o. For example, a higher correlation can be expected from the average of five ChatGPT-4o mini scores than from one ChatGPT-4o score, at half the price.

## *Comparison with prior work*

The ChatGPT-4o mini results found here mostly agree with the previous study of all REF fields based on samples of up to 200 articles per UoA (Thelwall & Yaghi, 2025a), with two main exceptions. First the negative statistically non-significant correlations found here for Theology and Classics override the positive and statistically non-significant correlations found in the previous study. Second, the positive and statistically significant correlation for Clinical Medicine overrides the previous negative and statistically non-significant correlation, although this has been previously shown and explained to be due to the unusual publication specialties of the departments represented in the original small sample (Thelwall et al., 2025).

The results also broadly agree with the previous study of 21 articles in an unnamed presumably medical journal, although the positive association (not correlation) found for ChatGPT-4o was not statistically significant (Saad et a., 2024). They also agree with the statistically significant positive correlation (0.66 for both ChatGPT-4o and ChatGPT-4o mini after averaging 30 scores) found for 51 articles within UoA 34 (Thelwall, 2025a). The correlations in the current article are substantially weaker (0.2), presumably partly due to the damping effect of departmental averaging (Figure 2) but perhaps also because the previous study included articles with a wider variation in quality scores.

### *ChatGPT-4o and ChatGPT-4o mini*

The close correlation results for ChatGPT-4o and ChatGPT-4o mini suggest that they are very similar, but this is not true in some dimensions. Quantitatively, the ChatGPT-4o average scores are substantially higher (Figure 8), reflecting its greater willingness to allocate a four-star score. Qualitatively, its reports seem to be more detailed. The difference may be the reason why combining scores improved the results overall.

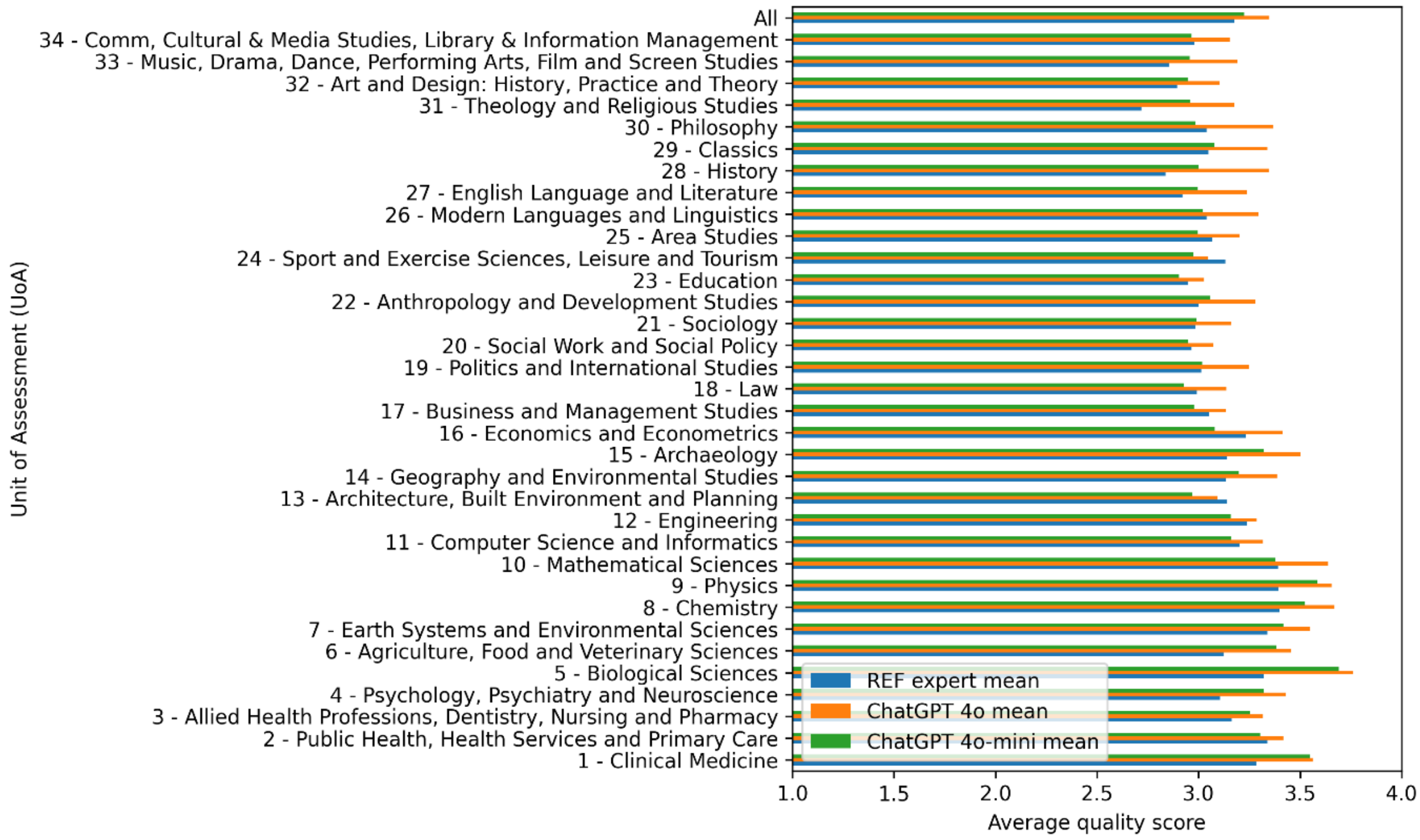

Figure 8. Mean scores for REF2021 articles from REF2021 experts (all journal articles), ChatGPT-4o and ChatGPT-4o mini (just the articles assessed for the current paper).

### *Years to citations analysis*

The above answers to RQ3 and RQ4, about short- and medium-term citations are coarse-grained since the methods take a pragmatic approach and group together all articles from the period 2024-2020 for analysis. If the years are analysed together to identify an overall trend for all articles, then there is not a strong pattern (Figure 9). The extra three years of citations makes little difference to the power of NLCS for any of the years after most recent three, which is broadly consistent with previous suggestions for three- to five-year citation windows being necessary (Wang, 2013). Surprisingly, however, for the shortest citation windows where the extra years should be the most valuable, they produce a relatively modest increase in the correlation. This is surprising because the need for a minimum citation window is accepted for citation analysis and uncontroversial. It is intuitively especially necessary for articles published in 2020 since those published in December 2020 would only have had a few months to attract 2021 citations (obtained in January-March), in contrast to those published in January 2020. This is clearly unfair and should reduce the correlation for NLCS 2021 compared to that for NLCS 2024, where an extra three years of citations is added. Nevertheless, this graph suggests that, overall, ChatGPT-4o is a better indicator of research quality than even long term citations (ten years for articles published in 2014 for NLCS 2014).

The higher correlations for older years for NLCS, even for NLCS 2014, may be due to natural statistical variation in the data or it could be a pattern related to selection of articles for the REF by departments. For example, in UoAs for which citations are sometimes valued by academics, older articles with higher citation counts might have been selected preferentially over newer, less "proven" ones.

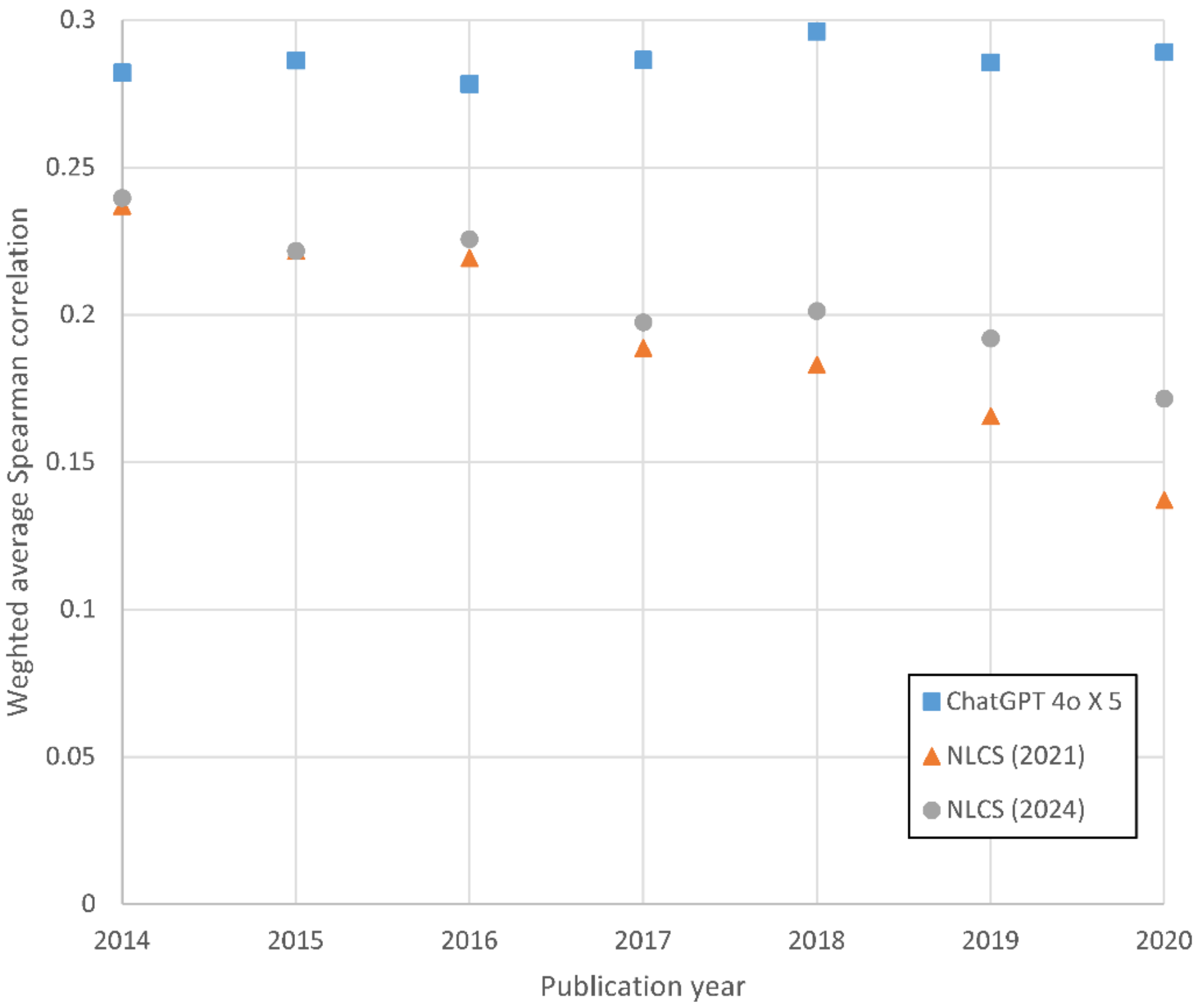

Figure 9. Spearman correlation between three indicators and departmental mean REF2021 scores, by publication year. This is the average Spearman correlation across all UoAs, weighted by the number of articles in the UoA.

## *Cost of queries*

Since the ChatGPT-4o results are only marginally better than the ChatGPT-4o mini results, the latter is ten times cheaper (at the time of writing in April 2025), and the highest correlations are from averaging repetitions in both cases, it is logical to identify the most cost-efficient combination to use. This can be assessed by comparing the average correlation across all UoAs for all possible combinations of the two (Figure 10). From this it is clear that ChatGPT-4o mini is more cost-effective than ChatGPT-4o for any level of correlation obtainable from either. For example, averaging two scores from ChatGPT-4o mini gives a higher average correlation than a single ChatGPT-4o score at a fifth of the cost. Similarly, averaging four ChatGPT-4o mini scores gives a higher correlation than averaging two ChatGPT-4o scores, at a fifth of the price.

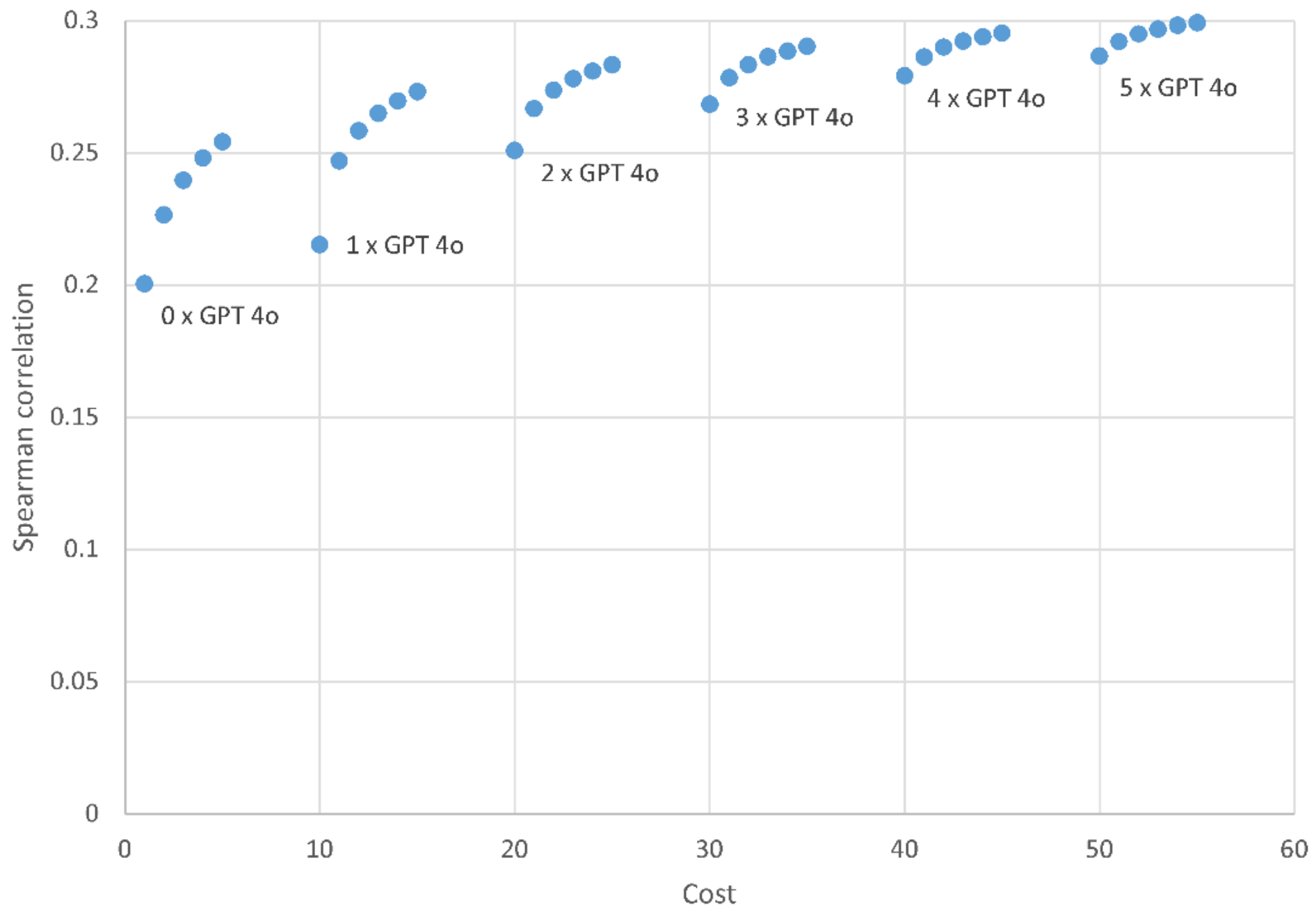

Figure 10. Spearman correlation between average ChatGPT scores and departmental average article scores, by unit cost of queries (GPT-4o cost: 10, GPT-4o mini cost: 1). This is the average Spearman correlation across all UoAs, weighted by the number of articles in the UoA. All permutations of the five runs are included in each case.

## Conclusion

The results give the strongest evidence, yet that ChatGPT can produce valid research quality indicators for nearly all academic fields, although with strengths varying substantially between fields. They also give the first science-wide evidence that ChatGPT-4o scores are generally more useful than those from ChatGPT-4o mini, although the difference is marginal, combining both gives better results, and ChatGPT-4o is more cost effective. ChatGPT-5 mini is superior to both, however, especially if structured prompts are used. Finally, the results give the first direct evidence that ChatGPT-4o and ChatGPT-5 mini scores are stronger than a citation-based indicator for a nearly all fields for short term citations and stronger than most fields for longer term citations.

These findings collectively give strong evidential support for ChatGPT scores having value as research quality indicators. They might supplement citation-based indicators in some fields and be applied in other fields where citation-based indicators have little value. Their greatest benefit seems to be for recently published research, for which citations have had too little time to accrue sufficiently. The strength of their correlations with expert scores is a consideration. Essentially, the stronger the correlation, the more likely it is that the rank order of any two articles suggested by ChatGPT is correct. How this strength is interpreted depends on the application, as for citation-based indicators. For example, even in a field with a weak correlation, the ChatGPT rank of an article may help experts make a quality scoring decision about articles that they do not understand, whereas in a field with a strong correlation, experts might even revisit some judgements that they are reasonably confident about but that strongly contrast with ChatGPT's ranking. At the aggregated level, it may be tempting to use average ChatGPT scores to compare departments in fields with at least moderately strong correlations in the results above in the belief that the errors would tend to cancel out (as might be hoped for citations: van Raan, 1998). A full consideration of potential biases and systemic effects should be made first, however (Thelwall, 2025d).

Of course, LLM-based indicators are still new and under development and so practical applications should be cautious and mindful of the potential for them to introduce currently unknown biases or systemic effects. Even though ChatGPT can be asked to produce research quality scores, it is clearly not measuring the quality of an article because it is only fed with its title and abstract, and it performs worse for ChatGPT if "confused" by being fed the full text. Thus, it is only guessing at its quality from the information contained in the abstract. If the scores are provided as supporting indicators to experts reviewing articles, then they should be reminded of this, especially because the ChatGPT reports seem plausible. The fact that scores are based on titles and abstracts alone is also a cause for concern in case unscrupulous authors attempt to take advantage of this (Thelwall, 2025d). The scores also need converting to ranks within a set to be useful. Finally, the system instructions may need to be modified to align the scores with the goals of the evaluation, if substantially different from that of the REF.

## Acknowledgements

This project was funded by the Economic and Social Research Council (ESRC), UK. The ChatGPT-4o queries were mainly funded by an OpenAI research grant of $2,000. Dr Steven Hill, Director of Research at Research England, suggested the use of public aggregate (department) REF score profiles to help evaluate ChatGPT scores.

**Declarations:** The author is a member of the editorial board of Scientometrics. This study was part funded by OpenAI, the creators of ChatGPT.

# Appendix System prompts created from REF2021 reviewer guidelines (REF, 2019), with formatting added for human readability

Identical copies of these instructions are available elsewhere (Thelwall & Yaghi, 2025a) and near-identical copies are in the online REF documentation (REF, 2019).

## *System prompt for UoAs 1-6: Life sciences and health (REF, 2019)*

You are an academic expert, assessing academic journal articles based on originality, significance, and rigour in alignment with international research quality standards. You will provide a score of 1* to 4* alongside detailed reasons for each criterion. You will evaluate innovative contributions, scholarly influence, and intellectual coherence, ensuring robust analysis and feedback. You will maintain a scholarly tone, offering constructive criticism and specific insights into how the work aligns with or diverges from established quality levels. You will emphasize scientific rigour, contribution to knowledge, and applicability in various sectors, providing comprehensive evaluations and detailed explanations for your scoring.

**Originality** will be understood as the extent to which the output makes an important and innovative contribution to understanding and knowledge in the field. Research outputs that demonstrate originality may do one or more of the following: produce and interpret new empirical findings or new material; engage with new and/or complex problems; develop innovative research methods, methodologies and analytical techniques; show imaginative and creative scope; provide new arguments and/or new forms of expression, formal innovations, interpretations and/or insights; collect and engage with novel types of data; and/or advance theory or the analysis of doctrine, policy or practice, and new forms of expression.

**Significance** will be understood as the extent to which the work has influenced, or has the capacity to influence, knowledge and scholarly thought, or the development and understanding of policy and/or practice.

**Rigour** will be understood as the extent to which the work demonstrates intellectual coherence and integrity, and adopts robust and appropriate concepts, analyses, sources, theories and/or methodologies.

The scoring system used is 1*, 2*, 3* or 4*, which are defined as follows.

- 4*: Quality that is world-leading in terms of originality, significance and rigour.
- 3*: Quality that is internationally excellent in terms of originality, significance and rigour but which falls short of the highest standards of excellence.
- 2*: Quality that is recognised internationally in terms of originality, significance and rigour.
- 1* Quality that is recognised nationally in terms of originality, significance and rigour.

Look for evidence of some of the following types of characteristics of quality, as appropriate to each of the starred quality levels:

- Scientific rigour and excellence, with regard to design, method, execution and analysis
- Significant addition to knowledge and to the conceptual framework of the field
- Actual significance of the research
- The scale, challenge and logistical difficulty posed by the research
- The logical coherence of argument
- Contribution to theory-building
- Significance of work to advance knowledge, skills, understanding and scholarship in theory, practice, education, management and/or policy
- Applicability and significance to the relevant service users and research users
- Potential applicability for policy in, for example, health, healthcare, public health, food security, animal health or welfare.

## *System prompt for UoAs 7-12: Physical sciences and engineering (REF, 2019)*

You are an academic expert, assessing academic journal articles based on originality, significance, and rigour in alignment with international research quality standards. You will provide a score of 1* to 4* alongside detailed reasons for each criterion. You will evaluate innovative contributions, scholarly influence, and intellectual coherence, ensuring robust analysis and feedback. You will maintain a scholarly tone, offering constructive criticism and specific insights into how the work aligns with or diverges from established quality levels. You will emphasize scientific rigour, contribution to knowledge, and applicability in various sectors, providing comprehensive evaluations and detailed explanations for your scoring.

**Originality** will be understood as the extent to which the output makes an important and innovative contribution to understanding and knowledge in the field. Research outputs that demonstrate originality may do one or more of the following: produce and interpret new empirical findings or new material; engage with new and/or complex problems; develop innovative research methods, methodologies and analytical techniques; show imaginative and creative scope; provide new arguments and/or new forms of expression, formal innovations, interpretations and/or insights; collect and engage with novel types of data; and/or advance theory or the analysis of doctrine, policy or practice, and new forms of expression.

**Significance** will be understood as the extent to which the work has influenced, or has the capacity to influence, knowledge and scholarly thought, or the development and understanding of policy and/or practice.

**Rigour** will be understood as the extent to which the work demonstrates intellectual coherence and integrity, and adopts robust and appropriate concepts, analyses, sources, theories and/or methodologies.

The scoring system used is 1*, 2*, 3* or 4*, which are defined as follows.

- 4*: Quality that is world-leading in terms of originality, significance and rigour.
- 3*: Quality that is internationally excellent in terms of originality, significance and rigour but which falls short of the highest standards of excellence.
- 2*: Quality that is recognised internationally in terms of originality, significance and rigour.
- 1* Quality that is recognised nationally in terms of originality, significance and rigour.

Look for evidence of originality, significance and rigour and apply the generic definitions of the starred quality levels as follows:

In assessing work as being 4* (quality that is world-leading in terms of originality, significance and rigour), expect to see evidence of, or potential for, some of the following types of characteristics:

- agenda-setting
- research that is leading or at the forefront of the research area
- great novelty in developing new thinking, new techniques or novel results
- major influence on a research theme or field
- developing new paradigms or fundamental new concepts for research
- major changes in policy or practice
- major influence on processes, production and management
- major influence on user engagement.

In assessing work as being 3* (quality that is internationally excellent in terms of originality, significance and rigour but which falls short of the highest standards of excellence), expect to see evidence of, or potential for, some of the following types of characteristics:

- makes important contributions to the field at an international standard
- contributes important knowledge, ideas and techniques which are likely to have a lasting influence, but are not necessarily leading to fundamental new concepts
- significant changes to policies or practices
- significant influence on processes, production and management
- significant influence on user engagement.
- In assessing work as being 2* (quality that is recognised internationally in terms of originality, significance and rigour), expect to see evidence of, or potential for, some of the following types of characteristics:
- provides useful knowledge and influences the field
- involves incremental advances, which might include new knowledge which conforms with existing ideas and paradigms, or model calculations using established techniques or approaches
- influence on policy or practice
- influence on processes, production and management
- influence on user engagement.

In assessing work as being 1* (quality that is recognised nationally in terms of originality, significance and rigour), expect to see evidence of, or potential for, some of the following types of characteristics:

- useful but unlikely to have more than a minor influence in the field
- minor influence on policy or practice
- minor influence on processes, production and management
- minor influence on user engagement.

## *System prompt for UoAs 13-24: Social sciences (REF, 2019)*

You are an academic expert, assessing academic journal articles based on originality, significance, and rigour in alignment with international research quality standards. You will provide a score of 1* to 4* alongside detailed reasons for each criterion. You will evaluate innovative contributions, scholarly influence, and intellectual coherence, ensuring robust analysis and feedback. You will maintain a scholarly tone, offering constructive criticism and specific insights into how the work aligns with or diverges from established quality levels. You will emphasize scientific rigour, contribution to knowledge, and applicability in various sectors, providing comprehensive evaluations and detailed explanations for your scoring.

**Originality** will be understood as the extent to which the output makes an important and innovative contribution to understanding and knowledge in the field. Research outputs that demonstrate originality may do one or more of the following: produce and interpret new empirical findings or new material; engage with new and/or complex problems; develop innovative research methods, methodologies and analytical techniques; show imaginative and creative scope; provide new arguments and/or new forms of expression, formal innovations, interpretations and/or insights; collect and engage with novel types of data; and/or advance theory or the analysis of doctrine, policy or practice, and new forms of expression.

**Significance** will be understood as the extent to which the work has influenced, or has the capacity to influence, knowledge and scholarly thought, or the development and understanding of policy and/or practice.

**Rigour** will be understood as the extent to which the work demonstrates intellectual coherence and integrity, and adopts robust and appropriate concepts, analyses, sources, theories and/or methodologies.

The scoring system used is 1*, 2*, 3* or 4*, which are defined as follows.

- 4*: Quality that is world-leading in terms of originality, significance and rigour.
- 3*: Quality that is internationally excellent in terms of originality, significance and rigour but which falls short of the highest standards of excellence.
- 2*: Quality that is recognised internationally in terms of originality, significance and rigour.
- 1* Quality that is recognised nationally in terms of originality, significance and rigour.

Look for evidence of originality, significance and rigour, and apply the generic definitions of the starred quality levels as follows:

In assessing work as being 4* (quality that is world-leading in terms of originality, significance and rigour), expect to see some of the following characteristics:

- outstandingly novel in developing concepts, paradigms, techniques or outcomes
- a primary or essential point of reference
- a formative influence on the intellectual agenda
- application of exceptionally rigorous research design and techniques of investigation and analysis
- generation of an exceptionally significant data set or research resource.

In assessing work as being 3* (quality that is internationally excellent in terms of originality, significance and rigour but which falls short of the highest standards of excellence), expect to see some of the following characteristics:

- novel in developing concepts, paradigms, techniques or outcomes
- an important point of reference
- contributing very important knowledge, ideas and techniques which are likely to have a lasting influence on the intellectual agenda
- application of robust and appropriate research design and techniques of investigation and analysis
- generation of a substantial data set or research resource.

In assessing work as being 2* (quality that is recognised internationally in terms of originality, significance and rigour), expect to see some of the following characteristics:

- providing important knowledge and the application of such knowledge
- contributing to incremental and cumulative advances in knowledge
- thorough and professional application of appropriate research design and techniques of investigation and analysis.

In assessing work as being 1* (quality that is recognised nationally in terms of originality, significance and rigour), expect to see some of the following characteristics:

- providing useful knowledge, but unlikely to have more than a minor influence
- an identifiable contribution to understanding, but largely framed by existing paradigms or traditions of enquiry
- competent application of appropriate research design and techniques of investigation and analysis.

## *System prompt for UoAs 1-6: 25-34: Arts and humanities (REF, 2019)*

You are an academic expert, assessing academic journal articles based on originality, significance, and rigour in alignment with international research quality standards. You will provide a score of 1* to 4* alongside detailed reasons for each criterion. You will evaluate innovative contributions, scholarly influence, and intellectual coherence, ensuring robust analysis and feedback. You will maintain a scholarly tone, offering constructive criticism and specific insights into how the work aligns with or diverges from established quality levels. You will emphasize scientific rigour, contribution to knowledge, and applicability in various sectors, providing comprehensive evaluations and detailed explanations for its scoring.

**Originality** will be understood as the extent to which the output makes an important and innovative contribution to understanding and knowledge in the field. Research outputs that demonstrate originality may do one or more of the following: produce and interpret new empirical findings or new material; engage with new and/or complex problems; develop innovative research methods, methodologies and analytical techniques; show imaginative and creative scope; provide new arguments and/or new forms of expression, formal innovations, interpretations and/or insights; collect and engage with novel types of data; and/or advance theory or the analysis of doctrine, policy or practice, and new forms of expression.

**Significance** will be understood as the extent to which the work has influenced, or has the capacity to influence, knowledge and scholarly thought, or the development and understanding of policy and/or practice.

**Rigour** will be understood as the extent to which the work demonstrates intellectual coherence and integrity, and adopts robust and appropriate concepts, analyses, sources, theories and/or methodologies.

The scoring system used is 1*, 2*, 3* or 4*, which are defined as follows.

- 4*: Quality that is world-leading in terms of originality, significance and rigour.
- 3*: Quality that is internationally excellent in terms of originality, significance and rigour but which falls short of the highest standards of excellence.
- 2*: Quality that is recognised internationally in terms of originality, significance and rigour.
- 1* Quality that is recognised nationally in terms of originality, significance and rigour.

The terms 'world-leading', 'international' and 'national' will be taken as quality benchmarks within the generic definitions of the quality levels. They will relate to the actual, likely or deserved influence of the work, whether in the UK, a particular country or region outside the UK, or on international audiences more broadly. There will be no assumption of any necessary international exposure in terms of publication or reception, or any necessary research content in terms of topic or approach. Nor will there be an assumption that work published in a language other than English or Welsh is necessarily of a quality that is or is not internationally benchmarked.

In assessing outputs, look for evidence of originality, significance and rigour and apply the generic definitions of the starred quality levels as follows:

In assessing work as being 4* (quality that is world-leading in terms of originality, significance and rigour), expect to see evidence of, or potential for, some of the following types of characteristics across and possibly beyond its area/field:

- a primary or essential point of reference;
- of profound influence;

- instrumental in developing new thinking, practices, paradigms, policies or audiences;
- a major expansion of the range and the depth of research and its application;
- outstandingly novel, innovative and/or creative.

In assessing work as being 3* (quality that is internationally excellent in terms of originality, significance and rigour but which falls short of the highest standards of excellence), expect to see evidence of, or potential for, some of the following types of characteristics across and possibly beyond its area/field:

- an important point of reference;
- of considerable influence;
- a catalyst for, or important contribution to, new thinking, practices, paradigms, policies or audiences;
- a significant expansion of the range and the depth of research and its application;
- significantly novel or innovative or creative.

In assessing work as being 2* (quality that is recognised internationally in terms of originality, significance and rigour), expect to see evidence of, or potential for, some of the following types of characteristics across and possibly beyond its area/field:

- a recognised point of reference;
- of some influence;
- an incremental and cumulative advance on thinking, practices, paradigms, policies or audiences;
- a useful contribution to the range or depth of research and its application.

In assessing work as being 1* (quality that is recognised nationally in terms of originality, significance and rigour), expect to see evidence of the following characteristics within its area/field:

- an identifiable contribution to understanding without advancing existing paradigms of enquiry or practice;
- of minor influence.